\documentclass[aps,twocolumn,showpacs,preprintnumbers,amsmath,amssymb]{revtex4}
\usepackage{latexsym}
\usepackage{graphicx}
\usepackage[T1]{fontenc}

\newcommand{\equationname}[1]{Eq.~(#1)}

\begin{document}

\title{Generation of polarization-entangled photons\\ using type-II doubly periodically poled lithium niobate waveguides}

\author{K. Thyagarajan}
\affiliation{Department of Physics, IIT Delhi, New Delhi 110016, India}

\author{K. Sinha}
\affiliation{University of Maryland, College Park, Maryland 20742, USA}

\author{J. Lugani}
\affiliation{Department of Physics, IIT Delhi, New Delhi 110016, India}

\author{S. Ghosh}
\affiliation{Department of Physics, IIT Delhi, New Delhi 110016, India}

\author{A. Martin}
\affiliation{Laboratoire de Physique de la Matière Condensée, CNRS UMR 6622, Université de Nice-Sophia Antipolis, Parc Valrose, 06108 Nice Cedex 2, France}

\author{D. B. Ostrowsky}
\affiliation{Laboratoire de Physique de la Matière Condensée, CNRS UMR 6622, Université de Nice-Sophia Antipolis, Parc Valrose, 06108 Nice Cedex 2, France}

\author{O. Alibart}
\affiliation{Laboratoire de Physique de la Matière Condensée, CNRS UMR 6622, Université de Nice-Sophia Antipolis, Parc Valrose, 06108 Nice Cedex 2, France}

\author{S. Tanzilli}\email{sebastien.tanzilli@unice.fr}
\affiliation{Laboratoire de Physique de la Matière Condensée, CNRS UMR 6622, Université de Nice-Sophia Antipolis, Parc Valrose, 06108 Nice Cedex 2, France}

\begin{abstract}
In this paper, we address the issue of the generation of non-degenerate cross-polarization-entangled photon pairs using type-II periodically poled lithium niobate. We show that, by an appropriate engineering of the quasi-phase-matching grating, it is possible to simultaneously satisfy the conditions for two 
spontaneous parametric down-conversion processes, namely ordinary pump photon down-conversion to either extraordinary signal and ordinary idler paired photons, or to ordinary signal and extraordinary idler paired photons. In contrast to single type-II phase-matching, these two processes, when enabled together, can lead to the direct production of cross-polarization-entangled state for non degenerate signal and idler wavelengths.
Such a scheme should be of great interest in applications requiring polarization-entangled non degenerate paired photons with, for instance, one of the entangled photons at an appropriate wavelength being used for local operation or for quantum storage in an atomic ensemble, and the other one at the typical wavelength of 1550\,nm for propagation through an optical fiber.
\end{abstract}

\maketitle

\section{Introduction}
\label{intro}
Entangled photon pairs are attracting a lot of attention due to their varied applications in the field of quantum information science and technology. Spontaneous parametric down-conversion (SPDC) in $\chi ^{(2)}$ nonlinear crystals, such as lithium niobate (LN) or potassium titanyl phosphate (KTP), is one of the primary resources for generating polarization-entangled photon pairs~\cite{Kwiat_new_1995,Kwiat_ultrabright_1999,Fiorentino_Ultrabright_PolarEnt_without_2004,fedrizzi_wavelength-tunable_2007}. The particular use of waveguiding channel structures leads to enhanced non-linear efficiencies due to tight confinement of the interacting waves and long interaction lengths. Periodically poled lithium niobate (PPLN) is seen as an important substrate and lots of works taking advantage of the largest nonlinear coefficient, namely $d_{33}$ in proton exchanged PPLN wave\-guides, have been reported in literature for the generation of time-bin type entanglement~\cite{tanzilli_ppln_2002,thew_experimental_2002,honjo_generation_2007}. SPDC leads in this case to the generation of signal and idler photons having both extraordinary polarizations, therefore not exhibiting a direct polarization-entangled state.

To produce polarization entanglement, several schemes have been proposed in the past years. In Refs.~\cite{Fiorentino_Ultrabright_PolarEnt_without_2004,fedrizzi_wavelength-tunable_2007}, the authors took advantage of a single type-II periodically poled potassium titanyl phosphate (PPKTP) bulk crystal surrounded by an interferometric setup to create cross-polarized entangled photons emitted around the degenerate wavelength of 800\,nm. Other schemes are based on type-II, titanium in-diffused PPLN or Rb in-diffused PPKTP waveguides, to emit cross-polarized entangled photons at a degenerate telecom wavelength~\cite{suhara_generation_2007,fujii_bright_2007,martin_integrated_2009,zhong_high_2009}. An interesting technological solution comes from Kawashima and co-workers who employed a type-I down-conversion in a single-substrate of titanium in-diffused PPLN waveguide, having a half-wave plate inserted in the middle along its propagation axis, to generate same-polarization-entangled photons at 1550\,nm~\cite{kawashima_type_I_2009}. However in this experimental method, the poor quality of entanglement reported can essentially be attributed to the number of technical steps necessary to fabricate the photon pair generator.

A shared drawback in the latter schemes, in which the photons are emitted at the same wavelength, is that polarization entanglement can only be available provided they are correctly separated at a non-deterministic 50:50 beam-splitter.
Avoiding the beam-splitter issue amounts to creating non degenerate pairs of photons which have in addition interesting features for specific quantum communication and networking applications. More precisely, having one of the entangled photons in the visible band where atomic ensembles operate, and its twin at a telecom wavelength, opens the possibility of quantum storage for the qubit carried by the shortest wavelength photon~\cite{chaneliere_storage_2005,de_riedmatten_solid-state_2008}, while the other one can propagate through an optical fiber over long distances. In this frame, König and co-workers employed a interferometric set-up surrounding a type-I bulk PPLN crystal, to emit non degenerate (795 and 1609\,nm) cross-polarized entangled photons~\cite{Konig_Spectrally_Bright_Polar_Source_2005}. In addition, Jiang and Tomita used two separated type-I PPLN waveguides mounted in a fiber loop to generate polarization-entangled photons at the non degenerate wavelengths of 1434 and 1606\,nm~\cite{Jiang_PolarEnt_PPLNW_fiberloop_2007}. In the latter case, each PPLN waveguide is responsible for the generation of one of the two components of the desired entangled state.
Note that most of the abovementioned solutions are outlined and tested experimentally in Ref.~\cite{fiorentino_compact_2008}.
Although the reported results are of prime interest, all of these solutions are based on interferometric setups requiring stringent control and stabilization.

In this paper, we address the issue of generating non-degenerate, cross-polarized, entangled photon pairs directly from a type-II, titanium in-diffused, PPLN waveguide. We show that this is made possible by an appropriate engineering of the quasi-phase-matching (QPM) grating so as to satisfy two SPDC processes simultaneously, namely down-conversion of ordinary pump photon to either extraordinary signal and ordinary idler paired photons or conversely. In contrast to standard type-II phase-matching based on a single process either in a bulk or waveguide configuration~\cite{Fiorentino_Ultrabright_PolarEnt_without_2004,fedrizzi_wavelength-tunable_2007,suhara_generation_2007,fujii_bright_2007,martin_integrated_2009,zhong_high_2009}, these two processes, when enabled together, can lead to the direct production of cross-polarization-entangled state for non degenerate signal and idler wavelengths without the need for an interferometric setup, or a 50:50 beam-splitter, as discussed above. In this case, a simple wavelength demultiplexer can be used to separate the twins. As already mentioned, such a scheme should be of great interest in applications requiring non-degenerate polarization-entangled photons, with for instance one of them adapted to local quantum operations in the visible band, and the other suitable for propagation through standard telecom optical networks.

In Sec.~\ref{Sec_2_QM_SPCD} we give a brief account of the basic quantum mechanical analysis leading to the generation of polarization-entangled photon pairs from such two coupled interaction processes. We specifically investigate the calculation of the emission bandwidth.
In Sec.~\ref{Sec_3_Engin_QPM} we describe a technique to achieve simultaneous quasi-phase-matching condition for both the desired SPDC processes. In Sec.~\ref{Sec_4_num_simul} we present the results of numerical simulations addressing both the possibility to generate maximally entangled states and the practicality of the idea. We specifically discuss in details the different issues that can be encountered both in terms of bandwidth and brightness. Finally, we give a brief conclusion.

\section{Quantum mechanical analysis of SPDC for two simultaneous QPM conditions}
\label{Sec_2_QM_SPCD}
We consider the process of parametric down-conversion in a titanium in-diffused waveguide in $z$-cut, $x$-propagating lithium niobate and assume that the substrate is poled such that two separate QPM conditions can be satisfied simultaneously. We also assume that the nonlinear interaction takes place between the fundamental guided modes of the waveguide. The pump at a wavelength $\lambda _{p}$ is assumed to be $y$-polarized corresponding to an ordinary polarization and the two SPDC processes lead to the generation of signals and idlers with both ordinary and extraordinary polarization components. For the considered pump powers, the pump field can be assumed, as usual, to be a non depleted classical field ($E_{p0}$), while signal and idler fields are considered to be represented by quantum operators. 
Thus, the electric field distributions for the pump, signal (ordinary and extraordinary polarizations), and idler (ordinary and extraordinary polarizations) modes are given by

\begin{widetext}
\begin{eqnarray}
\text{Pump (o): } & \vec {E}_{po} =\dfrac{1}{2}e_{po} \left( {\vec {r}} \right)E_{p0} \left( 
{e^{i\left( {k_p x-\omega_p t} \right)}+e^{-i\left( {k_p x-t} \right)}} \right)\hat{y}\label{Eq_01}\\
\text{Signal (o): } & \hat{E}_{so} =i\int d\omega_s e_{so} \left( {\vec {r}} \right)\sqrt{\dfrac{\hbar \omega_{s} }{2\varepsilon _{so} L_{int}}} \left( {\hat{a}_{so}e^{ik_{so} x}-\hat{a}_{so}^{\dagger} e^{-ik_{so} x}} \right)\hat{y}\label{Eq_02}\\
\text{Signal (e): } & \hat{E}_{se} =i\int d\omega_s e_{se} \left( {\vec {r}} \right)\sqrt  {\dfrac{\hbar \omega_{s} }{2\varepsilon _{se} L_{int}}} \left( {\hat{a}_{se} e^{ik_{se} x}-\hat{a}_{se}^{\dagger} e^{-ik_{se} x}} \right)\hat{z}\label{Eq_03}\\
\text{Idler (o): } & \hat{E}_{io} =i\int d\omega_i e_{io} \left( {\vec {r}} \right)\sqrt{\dfrac{\hbar \omega_{i} }{2\varepsilon _{io} L_{int}}} \left( {\hat{a}_{io}e^{ik_{io} x}-\hat{a}_{io}^{\dagger} e^{-ik_{io} x}} \right)\hat{y}\label{Eq_04}\\
\text{Idler (e): } &\hat{E}_{ie} =i\int d\omega_i e_{ie} \left( {\vec {r}} \right)\sqrt {\dfrac{\hbar \omega_{i} }{2\varepsilon _{ie} L_{int}}} \left( {\hat{a}_{ie} e^{ik_{ie} x}-\hat{a}_{ie}^{\dagger} e^{-ik_{ie} x}} \right)\hat{z}\label{Eq_05},
\end{eqnarray} 
\end{widetext}
where the first subscript refers to whether it corresponds to pump (p), signal (s) or idler (i), the second subscript refers to whether the wave is ordinary (o) or extraordinary (e), $e_{po}(\overrightarrow{r})$, $e_{so}(\overrightarrow{r})$, $e_{io}(\overrightarrow{r})$ represent the transverse dependence of the modal fields corresponding to the ordinary waves at pump, signal and idler respectively,
$e_{se}(\overrightarrow{r})$ and $e_{ie}(\overrightarrow{r})$ represent the transverse dependence of the modal fields of the extraordinary polarization at the signal and idler wavelengths, $L_{int}$ represents the interaction length, and $\epsilon_{p,q}\,(p=\{s,i\};\,q=\{o,e\})$ corresponds to the optical dielectric permittivity of lithium niobate.

Assuming that the signal and idler frequencies generated by the two SPDC processes to be the same, the energy conservation equation is given by:
\begin{equation}
\omega_p =\omega_{s} +\omega_{i}\label{Eq_06}.
\end{equation}

Quasi phase-matching conditions for the two processes require the corresponding spatial frequencies to be given by
\begin{eqnarray}
K_1=\dfrac{2\pi}{\Lambda_1}=2\pi \left(\dfrac{n_{po}}{\lambda_p}-\dfrac{n_{so}}{\lambda_{s}}-\dfrac{n_{ie}}{\lambda_{i}}\right)\label{Eq_08}\\
K_2=\dfrac{2\pi}{\Lambda_2}=2\pi \left(\dfrac{n_{po}}{\lambda_p}-\dfrac{n_{se}}{\lambda_{s}}-\dfrac{n_{io}}{\lambda_{i}}\right)\label{Eq_09},
\end{eqnarray}
where $\lambda_s$ and $\lambda_i$ represent free space signal and idler wavelengths satisfying the phase-matching conditions for both the processes.

\begin{figure}[h!]
\resizebox{1\columnwidth}{!}{\includegraphics{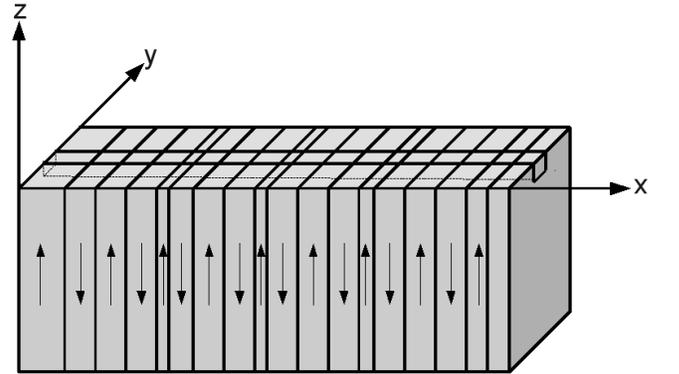}}
\caption{Waveguide geometry for generating polarization-entangled pair of photons using simultaneous SPDC to $y$-polarized signal, $z$-polarized idler and $z$-polarized signal, $y$-polarized idler. The lithium niobate substrate has its optic axis along the $z$-direction and is assumed to be poled according to the functional dependence $f_{1}(x) \, f_{2}(x)$.}
\label{Fig_1}
\end{figure}

The second order non-linear polarization generated in the medium is given by
\begin{equation}
P_i^{NL} =2\varepsilon _0 \sum\limits_{j,k} {d_{ijk} E_j E_k},
\label{Eq_10}
\end{equation}
where $E_{j}$ represents the $j^{th}$ component of the total electric field within the medium. For the case under consideration, the components of the total electric field are,
\begin{equation}
\begin{array}{l}
E_1 = 0 \\ 
E_2 = E_{po} +E_{so} +E_{io} \\ 
E_3 = E_{se} +E_{ie},
\end{array}
\label{Eq_11}
\end{equation}
where indices 1, 2 and 3 stand for $x$-, $y$- and $z$-components as depicted in \figurename{~\ref{Fig_1}}.

Using a noise-free model and the expressions for the electric fields in \equationname{\ref{Eq_10}} and using the rotating wave approximation and energy conservation, we obtain the following expression for the interaction Hamiltonian:
\begin{widetext}
\begin{equation}
\begin{array}{l c l}
\hat{H}_{int} &=&\int d\omega_s\left( {\dfrac{ E_{p0} \hbar \sqrt {\omega_s \omega_i } }{{L_{int}}}} \right) \int\limits_0^{L_{int}} { d_{24} \left[ {\left( {\dfrac{I_{oe} }{n_{so} n_{ie} }} \right)\left( {\hat{a}_{so}^{\dagger} \hat{a}_{ie}^{\dagger} e^{i\left( {\left( {k_p - k_{so} - k_{ie} } \right)x-\omega_p t} \right)}+\hat{a}_{so} \hat{a}_{ie} e^{-i\left( {\left( {k_p -k_{so} -k_{ie} } \right)x-\omega_p t} \right)}} \right)} \right.}\\
&+&\left.\left( {\dfrac{I_{eo} }{n_{se} n_{io} }} \right)\left( {\hat{a}_{se}^{\dagger} \hat{a}_{io}^{\dagger} e^{i\left( {\left( {k_p -k_{se} -k_{io} } \right)x-\omega_p t} \right)}+\hat{a}_{se} \hat{a}_{io} e^{-i\left( {\left( {k_p -k_{se} -k_{io} } \right)x-\omega_p t} \right)}} \right)\right] dx,
\end{array}
\label{Eq_21}
\end{equation}
where
\begin{equation}
\begin{array}{l}
I_{oe} =\int\!\!\!\int {e_{po} \left( {\vec {r}} \right)e_{so} \left( {\vec 
{r}} \right)e_{ie} \left( {\vec {r}} \right)dydz} \\ 
I_{eo} =\int\!\!\!\int {e_{po} \left( {\vec {r}} \right)e_{se} \left( {\vec 
{r}} \right)e_{io} \left( {\vec {r}} \right)dydz} \\ 
\end{array}
\label{Eq_20}
\end{equation}
denote the overlap integrals between the pump, signal and idler over the transverse coordinates. 

We show in Sec.~\ref{Sec_3_Engin_QPM} that it is possible to have two independent spatial frequency components in the nonlinear coefficient variation along the propagation direction. The effective nonlinear coefficient $\overline{d}$ including the effect of periodic domain reversal is given as (see \equationname{\ref{Eq_44}} and the corresponding derivation for more details)
\begin{equation}
\begin{array}{l r l}
\bar{d} &=& d_{24} f_1 \left( x \right)f_2 \left( x \right)\\
&=& - \dfrac{4d_{24}}{\pi ^2}\left( {e^{iK_1 x}+e^{-iK_1 x}-e^{iK_2 x}-e^{-iK_2 x}}\right) +\, \mbox{terms at other spatial frequencies}.
\end{array}
\label{Eq_22}
\end{equation}
Replacing the nonlinear coefficient $d_{24}$ by $\overline{d}$ in \equationname{\ref{Eq_21}}, we obtain
\begin{equation}
\begin{array}{l c l}
\hat{H}_{int} &=&-\int d\omega_s\left( \dfrac{4d_{24} E_{p0} \hbar \sqrt{\omega_s\omega_i}}{\pi^2 L_{int}} \right)
\times\int\limits_0^{L_{int}} \left[ \dfrac{I_{oe}}{n_{so}n_{ie}} \left( \hat{a}_{so}^\dagger \hat{a}_{ie}^{\dagger} e^{-i(\omega_p t+\Delta {k_{oe}} x)}+\hat{a}_{so} \hat{a}_{ie} e^{i(\omega_p t + i\Delta {k_{oe}} x)} \right)\right.\\
&& \hspace{5cm} \left. +\dfrac{I_{eo} }{n_{se} n_{io} }\left( {\hat{a}_{se}^{\dagger} \hat{a}_{io}^{\dagger} e^{-i(\omega_p t + \Delta {k_{eo}} x)}} +\hat{a}_{se} \hat{a}_{io} e^{i(\omega_p t + \Delta {k_{eo}} x)} \right) \right] dx\\
&=&-\int d\omega_s\left(\dfrac{4d_{24} E_{p0} \hbar \sqrt {\omega_s \omega_i } }{\pi^2}\right)\times\Bigg[ \dfrac{I_{oe}}{n_{so}n_{ie}}\left( {\hat{a}_{so}^{\dagger}
\hat{a}_{ie}^{\dagger} e^{-i\omega_p t}+\hat{a}_{so} \hat{a}_{ie} e^{i\omega_p t} e^{i\Delta {k_{oe}}L_{int}}} \right)
e^{\frac{-i\Delta {k_{oe}}L_{int}}{2}} \mbox{sinc} \left(\dfrac{\Delta {k_{oe}}L_{int}}{2} \right) \\
&& \hspace{5cm} + \dfrac{I_{eo}}{n_{se}n_{io}}\left({\hat{a}_{se}^{\dagger} \hat{a}_{io}^{\dagger} e^{-i\omega_p t}+\hat{a}_{se} \hat{a}_{io} e^{i\omega_p t} e^{i\Delta {k_{eo}}L_{int}}} \right)e^{\frac{-i\Delta {k_{eo}}L_{int}}{2}} \mbox{sinc}\left(\dfrac{\Delta {k_{eo}}L_{int}}{2} \right) \Bigg]\\
&=&\int d\omega_s C_{oe}^{(1)} \left( {\hat{a}_{so}^{\dagger} \hat{a}_{ie}^{\dagger} e^{-i\omega_p t}+\hat{a}_{so} \hat{a}_{ie} e^{i\omega_p t}} \right)+C_{eo}^{(1)} \left( {\hat{a}_{so}^{\dagger} \hat{a}_{ie}^{\dagger} e^{-i\omega_p t}+\hat{a}_{so} \hat{a}_{ie} 
e^{i\omega_p t}} \right),
\end{array}
\label{Eq_23}
\end{equation}
where
\begin{equation}
\begin{array}{l}
C_{oe}^{(1)} =-\left( {\dfrac{4d_{24} E_{p0} \hbar \sqrt {\omega_s \omega_i } I_{oe} }{\pi ^2n_{so} n_{ie} }} \right)e^{\frac{-i\Delta {k_{oe}}L_{int}}{2}}  \mbox{sinc}\left(\Delta{k_{oe}} \dfrac{L_{int}}{2}\right) \\ 
C_{eo}^{(1)} =-\left( {\dfrac{4d_{24} E_{p0} \hbar \sqrt {\omega_s \omega_i } I_{eo} }{\pi ^2n_{se} n_{io} }} \right)e^{\frac{-i\Delta {k_{eo}}L_{int}}{2}} \mbox{sinc}\left(\Delta{k_{eo}} \dfrac{L_{int}}{2}\right).
\end{array}
\label{Eq_24}
\end{equation}
\end{widetext}
with
\begin{eqnarray}
\Delta k_{oe}&=&\frac{2\pi}{\Lambda_1}-2\pi \left(\frac{n_{po}}{\lambda_p}-\frac{n_{so}}{\lambda_{s}}-\frac{n_{ie}}{\lambda_{i}}\right)\\
\Delta k_{eo}&=&\frac{2\pi}{\Lambda_2}-2\pi \left(\frac{n_{po}}{\lambda_p}-\frac{n_{se}}{\lambda_{s}}-\frac{n_{io}}{\lambda_{i}}\right).
\end{eqnarray}

In writing \equationname{\ref{Eq_23}} we have only kept the terms that are close to the phase-matching.

Using the expression for the interaction Hamiltonian, we can use the interaction picture to obtain the output state as
\begin{equation}
\left| \Psi \right\rangle =i\int d\omega_s\left( {C_{oe} \left| {s_o ,i_e } \right\rangle 
+C_{eo} \left| {s_e ,i_o } \right\rangle } \right),
\label{Eq_37}
\end{equation}
where 
\begin{equation}
\begin{array}{l}
C_{oe} =-\dfrac{tC_{oe}^{(1)} }{\hbar } \\ 
C_{eo} =-\dfrac{tC_{eo}^{(1)} }{\hbar }. 
\end{array}
\label{Eq_38}
\end{equation}
Since $C_{oe}$ and $C_{eo}$ depend on $\omega_s$ (and hence $\omega_i$) through the sinc functions (see~\equationname{\ref{Eq_24}}), the output state will be entangled in the region of their overlap~\cite{saleh_2009}. The relative values of $C_{oe}$ and $C_{eo}$ will determine if the output state is maximally entangled or not. These are related to the overlap integrals and the ordinary and extraordinary effective indices of the interacting modes at the pump, signal and idler frequencies. In Sec.~\ref{Sec_4_num_simul}, we use practical values of various parameters and show that by properly determining the waveguide design, it is possible to obtain a maximally entangled state. 

The sinc functions in \equationname{\ref{Eq_24}} will determine the bandwidth of the two down-conversion processes, which in turn will be given by the wavelength variation of $\Delta k_{oe}$ and $\Delta k_{eo}$ and the interaction length $L_{int}$. In fact, by making Taylor series expansion of the effective indices around the central wavelengths, we can obtain the following approximate expressions for the signal bandwidths of the two processes:
\begin{equation}
\begin{array}{l}
\Delta \lambda_{oe} =\frac{\lambda_{s}^{2}}{L (N_{ie}-N_{so})} \\ 
\Delta \lambda_{eo} =\frac{\lambda_{s}^{2}}{L (N_{io}-N_{se})} ,
\end{array}
\label{Eq_36}
\end{equation}
where $N_{so}$, $N_{se}$, $N_{io}$ and $N_{ie}$ represent the group effective indices of the modes corresponding to the ordinary and extraordinary signal wavelengths, and to ordinary and extraordinary idler wavelengths, respectively, evaluated at the phase-matching wavelengths.
As we will see in Sec.\ref{Sec_4_num_simul}, the bandwidth ratio regarding the two processes will hence depend on the differences in group effective indices of the modes, which in turn will depend on the signal and idler wavelengths, the waveguide parameters, and the material.

\section{Engineering Quasi-Phase-Matching}
\label{Sec_3_Engin_QPM}

We saw in Sec.~\ref{Sec_2_QM_SPCD} that if the QPM conditions for the generation of down-converted pairs of extraordinary signal, ordinary idler and ordinary signal, extraordinary idler can be satisfied simultaneously, then it is possible to generate a polarization-entangled state. The spatial variation of the nonlinear coefficient along the propagation direction settles the available spatial frequencies. It is indeed possible to have both QPM conditions to be satisfied simultaneously by noting first that if a periodic function with fundamental spatial frequency $K_{0}$ is amplitude or phase modulated by another periodic function with spatial frequency $K_{p}\,(<K_{0})$, then the modulated function would have spatial frequency components at $nK_{0} +mK_{p}$ with $n=\pm 1, \pm2, \dots$ and $m=\pm 1, \pm2, \dots$ Since the amplitude of the Fourier coefficients determine the strength of the nonlinear interaction, only the lower order terms will have significant efficiency.

When poled, the nonlinear coefficient can be positive or negative (depending on the direction of the spontaneous polarization). We will therefore consider functions which have a given magnitude and are positive or negative. Let us assume that a periodic domain reversal with a period $\Lambda_{0}$ is 
modulated by another periodic reversal with a period $\Lambda_{p}\, (>\Lambda_{0})$. Thus the functional dependence of the nonlinear coefficient $d_{24}$ can be written as
\begin{equation}
\overline{d} = d_{24} f_1(x)f_2(x),
\label{Eq_39}
\end{equation}
where 
\begin{equation}
\begin{array}{l c c}
f_1(x)&=+1 ;& 0<x<\dfrac{\Lambda_0}{2}\\
&=-1;& \dfrac{\Lambda_0}{2}<x<\Lambda_0,
\end{array}
\label{Eq_40}
\end{equation}
with 
\begin{equation}
f_1(x+\Lambda_0)=f_1(x).
\label{Eq_41}
\end{equation}
And similarly,
\begin{equation}
\begin{array}{l c c}
f_2(x)&=+1 ;& 0<x<\dfrac{\Lambda_p}{2}\\
&=-1;& \dfrac{\Lambda_p}{2}<x<\Lambda_p,
\end{array}
\label{Eq_42}
\end{equation}
with 
\begin{equation}
f_2(x+\Lambda_p)=f_2(x).
\label{Eq_43}
\end{equation}

Simple Fourier series expansion of the functions $f_{1}(x)$ and $f_{2}(x)$ gives us the following expression for the expansion with only the terms corresponding to fundamental frequencies $K_{1}$ and $K_{2}$:
\begin{equation}
\overline{d}=-\dfrac{4d_{24}}{\pi^2} \left(e^{iK_1x}-e^{iK_2x}+e^{-iK_1x}-e^{-iK_2x}\right) ,
\label{Eq_44}
\end{equation}
where
\begin{equation}
K_1=K_0+K_p \text{ and } K_2=K_0-K_p.
\label{Eq_45}
\end{equation}

By choosing appropriate values of $K_{0}$ and $K_{p}$, it is thus possible to generate required spatial frequencies in the nonlinear variation so as to simultaneously quasi-phase-match the two different processes of down-conversion of ordinary pump to ordinary signal-extraordinary idler and extraordinary signal-ordinary idler pairs. 

\section{Numerical simulations and discussion}
\label{Sec_4_num_simul}
In this section we model a titanium in-diffused channel waveguide and use a scalar variational method (which is accurate enough to model weakly guiding waveguides) to obtain the effective indices and modal field profiles at the pump, signal and idler wavelengths. As an approximation we neglect the anisotropic nature of the waveguide and use the analysis to obtain the 
spatial frequencies required and the overlap integrals which will indicate whether it is possible to obtain a maximally entangled state or not. 

Titanium in-diffused channel waveguides can be described by a refractive index variation of the form~\cite{sharma_variational_1994}
\begin{equation}
\begin{array}{l c l l}
n^2(y,z)&=&n_b^2 +2n_b \Delta ne^{-y^2/w^2}e^{-z^2/h^2};&z<0 \\ 
&=&n_c^2;  &z>0,
\end{array}
\label{Eq_46}
\end{equation}
where the $z$ axis represents the optic axis, $n_{b}$ is the bulk substrate refractive index, $\Delta n$ is the maximum change in the refractive index due to titanium in-diffusion, $n_{c}$ is the refractive index of the air, $w$ is the width and $h$ is height-to-depth ratio of the waveguide. For the ordinary and extra ordinary polarizations we shall use $n_{bs}=n_{bo}$, $\Delta n=\Delta n_o$ and $n_{b}=n_{be}$, $\Delta n=\Delta n_e$, respectively.

Using this variational technique, we assume the trial modal field to be given by the following Hermite Gauss function based field~\cite{sharma_variational_1994,sharma_analysis_1993}:
\begin{equation}
\begin{array}{lcl c}
\psi _t(y,z)&=&\sqrt {\dfrac{16\alpha_y \alpha_z}{\pi wh}} \alpha_z 
\left( {\dfrac{z}{h}} \right)e^{-\alpha_y^2 y^2/w^2}e^{-\alpha_z^2 
z^2/h^2}; &z<0 \\ 
&=&0; &z>0,
\end{array}
\label{Eq_47}
\end{equation}
with $\alpha_{y}$ and $\alpha_{z}$ as the parameters to be determined through the maximization of the $n_{eff}$ value which is given by 
\begin{equation}
\begin{array}{l l}
n_{eff}^2 =& - \dfrac{1}{k_0^2 }\int\!\!\!\int {\left| {\nabla _T \psi(y,z)}\right|^2} dydz\\
& +\int\!\!\!\int {n^2} (y,z)\left| {\psi(y,z)} \right|^2dydz.
\end{array}
\label{Eq_48}
\end{equation}

Since the index difference between the lithium niobate substrate and air is high, we approximate the field in the cover to be zero. In principle, it is possible to estimate the field and the effective indices more accurately by using trial fields with more variational parameters. Using \equationname{\ref{Eq_47}}, \equationname{\ref{Eq_48}} gives
\begin{equation}
n_{eff}^2 = n_b^2 - \dfrac{\alpha_y^2h^2 + 3w^2\alpha_z^2}{k_0^2w^2h^2} + \dfrac{8n_b\Delta n\alpha_y\alpha_z^3}{(2\alpha_z^2+1)^{3/2}\sqrt{2\alpha_y^2+1}}.
\label{Eq_49}
\end{equation}
The above eigenvalue is maximized with respect to the variational parameters $\alpha_{y}$ and $\alpha_{z}$ to obtain the effective index and the field distributions. 
This formulation is used to obtain the modal fields and the effective indices of the propagating ordinary and extraordinary modes at pump, signal and idler wavelengths. 
The values of the lithium niobate substrate refractive index $n_{b}$ for different wavelengths and temperature were calculated using temperature dependent Sellemeir equation given in Ref.~\cite{Crystal_Tech}. Moreover, for the waveguide index variation, corresponding to ordinary and extraordinary waves, we have used the results given in Ref.~\cite{fouchet_wavelength_1987}.
In our case, the pump, signal and idler wavelengths are chosen to be 519, 780, and 1551\,nm. The QPM period required for ordinary signal and extraordinary idler can be deduced from \equationname{\ref{Eq_08}} and \equationname{\ref{Eq_09}}.

Analytical expressions for the overlap integrals can be derived from the pump, signal and idler electric field profiles, taken to be of the form given by \equationname{\ref{Eq_47}}. Substituting the latter in \equationname{\ref{Eq_24}} when the phase-matching conditions are exactly satisfied for the center wavelengths ($\Delta k_{oe}=0$ and $\Delta k_{eo}=0$), we obtain
\begin{widetext}
\begin{equation}
\dfrac{C_{oe}}{C_{eo}} = \dfrac{\sqrt{\alpha_{yso}}\alpha_{zso}^{3/2} \sqrt{\alpha_{yie}}\alpha_{zie}^{3/2} \sqrt{\alpha_{ypo}^2 + \alpha_{yse}^2 + \alpha_{yio}^2 } (\alpha_{zpo}^2 + \alpha_{zse}^2 + \alpha_{zio}^2)^2 n_{se}n_{io}}
{\sqrt{\alpha_{yse}}\alpha_{zse}^{3/2} \sqrt{\alpha_{yio}}\alpha_{zio}^{3/2} \sqrt{\alpha_{ypo}^2 + \alpha_{yxso}^2 + \alpha_{yie}^2 } (\alpha_{zpo}^2 + \alpha_{zso}^2 + \alpha_{zie}^2)^2 n_{so}n_{ie}}.
\label{Eq_52}
\end{equation} 
\end{widetext}

In the case of the generation of pure entangled states, i.e. without a noise counterpart that would induce a state description using a density operator, we quantify the output state to be maximally entangled or not by defining a parameter $\gamma$ as
\begin{equation}
\gamma = \dfrac{min(C_{oe},C_{eo})}{max(C_{oe},C_{eo})},
\label{Eq_53}
\end{equation}
which characterizes the relative probabilities to generate the two contributions to the state, i.e. ordinary signal and extraordinary idler photons, and conversely.
$\gamma$ ranges from 0 to 1, where 0 corresponds to a simple product state and 1 to a maximally entangled state.

\begin{figure}[h!]
\resizebox{1\columnwidth}{!}{\includegraphics{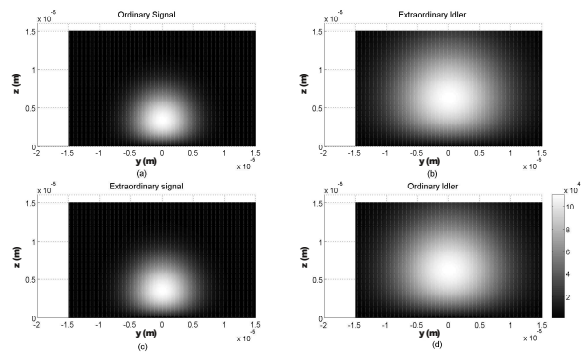}}
\caption{Field patterns corresponding to signal and idler wavelengths having ordinary and extraordinary polarizations.}
\label{Fig_patterns}
\end{figure}

In order to analyze the achievable degree of entanglement, we consider a titanium in-diffused waveguide with the specifications given in \tablename{~\ref{Table_1}}. These correspond to fabrication conditions under which the changes in the ordinary and extraordinary indices are almost equal~\cite{fouchet_wavelength_1987}. Simulations have been carried out for different values of $h$ and $w$. Figure~\ref{Fig_patterns} shows the transverse field patterns at the signal and idler wavelengths corresponding to ordinary and extraordinary polarizations for a waveguide design having $d=w=10\,\mu$m. The signal fields are more confined than the idler fields due to the shorter wavelength. The difference in field patterns leads to a decrease in the overlap integral; however, due to the shape of the fields, the quantities $C_{oe}$ and $C_{eo}$ (with $\Delta k_{oe}=0$ and $\Delta k_{eo}=0$) are made almost equal enabling the possibility to get a maximally entangled state.

\begin{figure}[h!]
\resizebox{1\columnwidth}{!}{\includegraphics{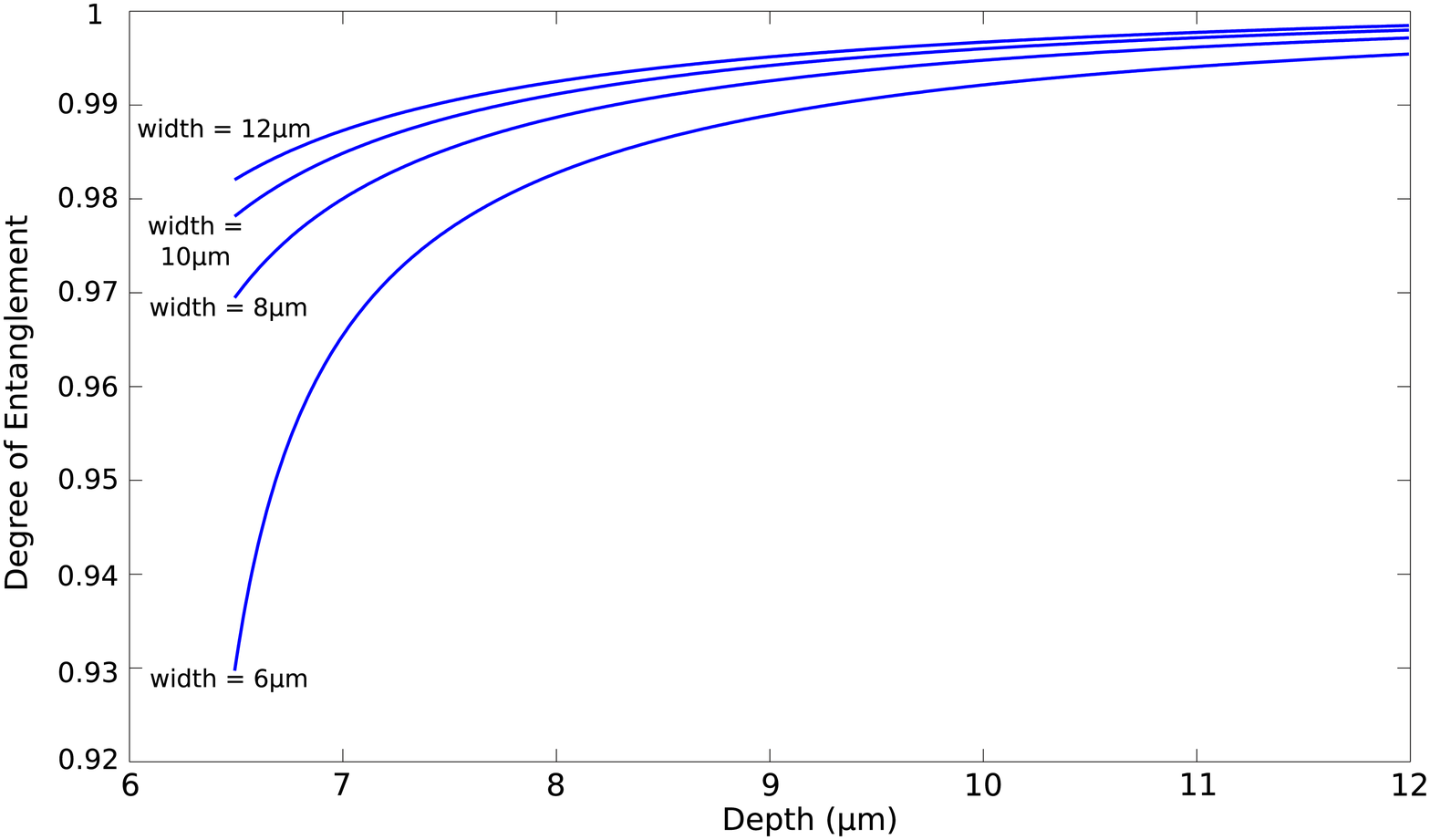}}
\caption{Variation of $\gamma$ as a function of the waveguide depth for different values of widths. Values of other parameters are given in \tablename{~\ref{Table_1}}.}
\label{Fig_depth}
\end{figure}

\begin{figure}[h!]
\resizebox{1\columnwidth}{!}{\includegraphics{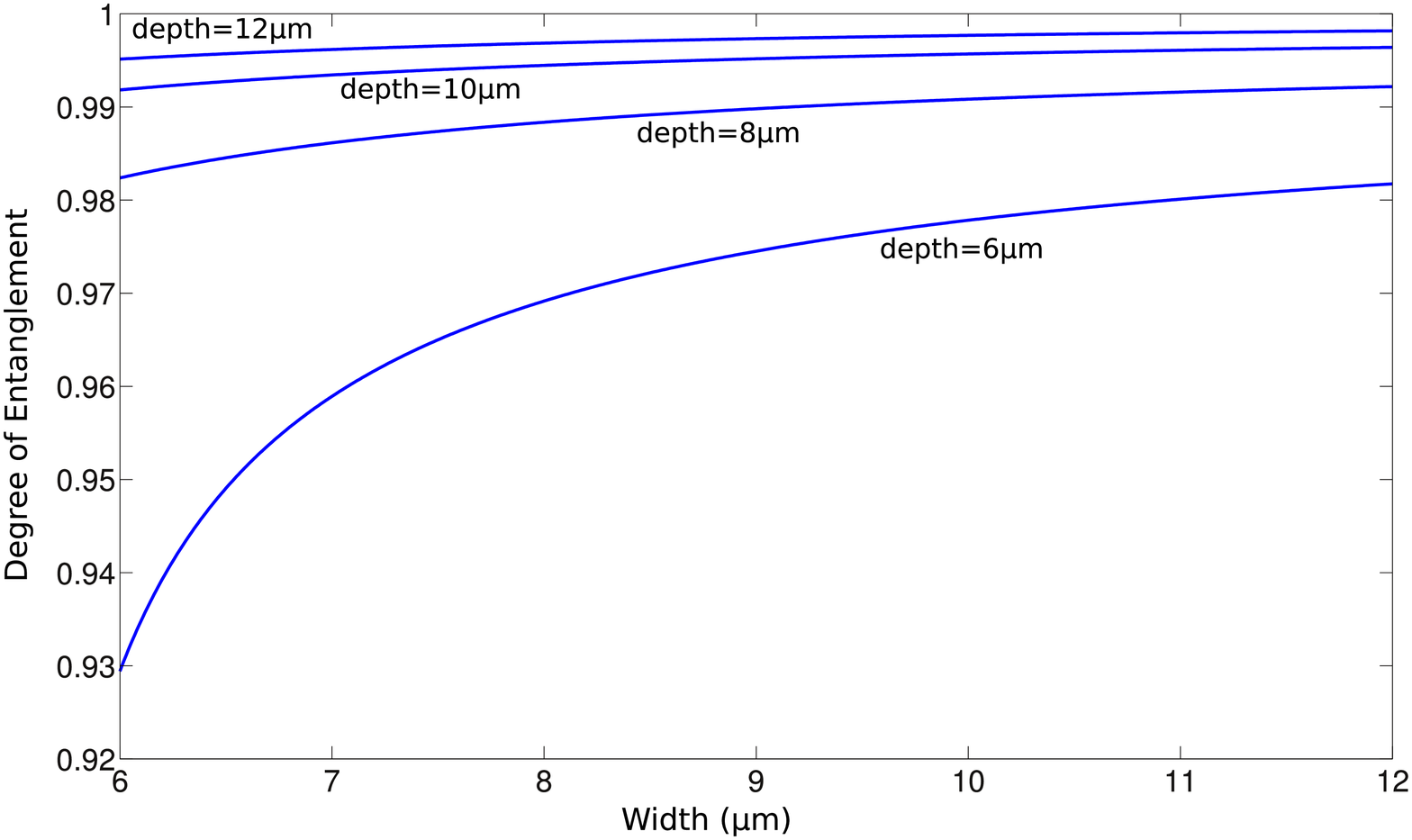}}
\caption{Variation of $\gamma$ as a function of the waveguide width for different values of depths. Values of other parameters are given in \tablename{~\ref{Table_1}}.}
\label{Fig_width}
\end{figure}

Figure~\ref{Fig_depth} shows the variation of $\gamma$ as a function of the waveguide depth for different waveguide widths and \figurename{~\ref{Fig_width}} shows the variation of $\gamma$ as a function of the waveguide width for different waveguide depths. It can be seen that maximally entangled states can be reached over a wide range of waveguide depths and widths for the chosen value of $\tau / d_{z}$ = 0.005.
\tablename{~\ref{Table_2}} shows the values of $\gamma$ and the spatial periods required to generate entangled pairs of photons. We see that for waveguide widths and depths of about 8\,$\mu $m and above, which are accessible from the technological side, the values of $\gamma$ are close to unity which correspond to obtaining maximally entanglement state.
For instance, when $d$ = 10\,$\mu $m and $w$ = 10\,$\mu $m, the required QPM periods are $\Lambda_{1}$ = 4.58 $\mu $m and $\Lambda_{2}$ = 3.65 $\mu $m for the corresponding value of $\gamma = 0.996$. These waveguide parameter values chosen to demonstrate the generation of almost maximally entangled states are consistent with technologically achievable values, as discussed in Ref.~\cite{martin_integrated_2009}.

In this context, the waveguide may no longer be single mode at the signal and idler wavelengths (780 and 1551\,nm), and of course at the pump wavelength (519\,nm), which is much shorter. However, it is possible to excite only the fundamental mode at the pump wavelength using segmented taper waveguides~\cite{Castaldini_tapers_2007}, and since the chosen phase-matching conditions only operates for fundamental modes from the pump field to signal and idler fields, the nonlinear process will automatically lead to the generation of photon pairs in the fundamental modes at signal and idler wavelengths.

\begin{figure}[h!]
\resizebox{1\columnwidth}{!}{\includegraphics{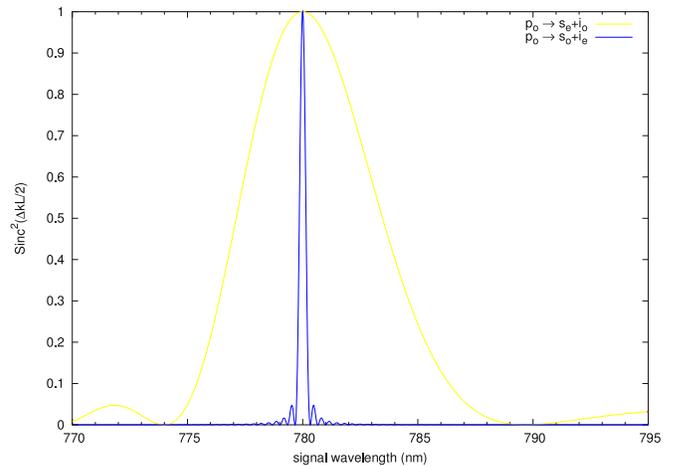}}
\caption{Calculated bandwidths for the two SPDC processes available from our QPM engineering at the signal wavelength of 780\,nm. The process leading to ordinary signal and extraordinary idler gives a bandwidth of 0.29\,nm associated with the blue (dark gray) curve, while the one leading to extraordinary signal and ordinary idler gives a bandwidth of 6.35\,nm associated with the yellow (light gray) curve.}
\label{Fig_bandwidth}
\end{figure}

Another condition for obtaining a maximally entangled state amounts to having identical bandwidths for the two enabled SPDC processes. As discussed in Sec.\ref{Sec_2_QM_SPCD}, it appears that the group effective indices experienced by signal and idler photons, whether they are ordinary or extraordinary polarized, are very different since their wavelengths are far from each other (780 and 1550\,nm, respectively). We naturally expect very different bandwidths for the two processes, as shown  in \figurename{~\ref{Fig_bandwidth}} which gives the normalized output spectra corresponding to the two considered processes, taken at the signal wavelength of 780\,nm. For the chosen waveguide parameters and interaction wavelengths, the two bandwidths are quite different leading to a ratio of about 22.
However, a maximally entangled state can be obtained by using a narrow bandpass filter to select only the region where the two sinc functions of \equationname{\ref{Eq_24}} are identical and equal to unity. The use of a filter having a bandpass shorter than the shortest bandwidth (here of 0.29\,nm) can erase the bandwidth distinguishability and let the coefficients $C_{eo}$ and $C_{oe}$ be the only relevant parameters concerning entanglement. In our case, recalling that the 780\,nm photons are dedicated to a lossy local quantum operation such as a quantum gate, a single filter can be used, placed for instance on the path of the 1550\,nm photons to avoid further losses. Here we can take advantage of very performant fiber Bragg grating filters developed in the frame of the telecommunications industry. This method, coupled to a coincidence detection technique, is commonly employed and known as non-local filtering.
Although in the example considered here the bandwidth difference between the two processes is large, we may mention that it should be possible to reduce the bandwidth difference advantageously by employing either another phase-matching interaction in terms of a different set of pump, signal and idler wavelengths, or another type of crystal such as PPKTP.

Note also that since signal and idler photons are orthogonally polarized, they will exit the waveguide at different times because of dispersion. The output state can be entangled by first separating the signal and idler photons using a wavelength division demultiplexer and then using birefringent crystals or a Michelson interferometer to compensate for the different velocities of the signal and idler photons~\cite{martin_integrated_2009}.

From the brightness side, the efficiency of the down-conversion processes are determined by the coefficients $C_{oe}$ and $C_{eo}$ and the additional bandpass filter discussed above. For a cm-long device, we expect a brightness on the order of $10^5$ pairs of photons created per second, per mW of pump power, and per GHz of bandwidth, as usually obtained with standard type II, titanium in-diffused PPLN waveguides~\cite{martin_integrated_2009}. This compares favorably to formerly reported solutions cited in Sec.~\ref{intro}~\cite{Fiorentino_Ultrabright_PolarEnt_without_2004,fedrizzi_wavelength-tunable_2007,takesue_generation_2005,suhara_generation_2007,fujii_bright_2007,martin_integrated_2009,zhong_high_2009,kawashima_type_I_2009}.
Regarding this, note that polarization-entangled states can also be obtained by having two separate grating sections on a single substrate, where the first section would phase-match one interaction while the second section would phase-match the other interaction. For comparison purpose, we assumed the same total length of substrate. On one hand, in the separate grating case, the interaction length for each down-conversion process would be reduced by a factor of 2 as compared to our configuration. On the other hand, the effective nonlinear coefficient in the case of two separate gratings would be higher by a factor of $\frac{2}{\pi}$. As a result, this leads to an overall efficiency for our compound grating scheme higher by a factor of $\left(\frac{4}{\pi}\right)^2 \simeq 1.6$, as compared to the separate grating configuration. In addition, one has to note that the bandwidth ratio between the two down-conversion processes would be the same since it is independent of the interaction length (see \equationname{\ref{Eq_36}}).

We may finally mention that the idea presented in this paper can also be applied to the case of modal entangled states generation for which two separate down-conversion processes have to be phase-matched simultaneously as depicted in Ref.~\cite{saleh_2009,silberhorn_2009}.

\section{Conclusion}
\label{Sec_conclusion}

In this paper we have addressed the issue of generation of polarization-entangled pairs using lithium niobate and show that by appropriately engineering the QPM grating it is possible to simultaneously satisfy the conditions for both SPDC processes namely ordinary pump photon down-conversion to either an extraordinary signal and ordinary idler photon pair or to an ordinary signal and extraordinary idler photon pair. This leads to a direct production of polarization-entangled state from the interaction 
process with non degenerate signal and idler wavelengths~\cite{suhara_quasi-phase_2009}. Such a scheme should be of great interest in applications requiring polarization-entangled non degenerate photon pairs with one of the entangled photons at an appropriate wavelength being used for interaction with an atomic system and the other at a typical wavelength of 1550 nm for propagation through an optical fiber.

\section*{Acknowledgement}
\label{Sec_Acknow}

The work reported in the paper was partially supported by the Department of Science and Technology, India and CNRS, France through an Indo French networking project. Their support is gratefully acknowledged. A. M. also acknowledges CNRS and the Regional Council PACA for their joint financial support.

\vspace{5cm}

\begin{table}[h!]
\begin{center}
\begin{tabular}{|l|l|l|}
\hline
$\lambda$ (nm)& 
$\Delta n_{o}$& 
$\Delta n_{e}$ \\
\hline
519& 
0.0038& 
0.0037 \\
\hline
780& 
0.0034& 
0.0030 \\
\hline
1550& 
0.0025& 
0.0025 \\
\hline
\end{tabular}
\end{center}
\caption{Values of $\Delta n_{o}$ and $\Delta n_{e}$ at the pump, signal and idler wavelengths considered in the numerical simulation.}
\label{Table_1}
\end{table}

\begin{table}[h!]
\begin{center}
\begin{tabular}{|l|l|l|l|l|}
\hline
Depth & Width & Degree of & $\Lambda _{1}$ & $\Lambda _{2}$\\ 
($\mu$m) & ($\mu$m)& entanglement & ($\mu$m) & ($\mu$m)\\
\hline
6.5& 
6.0& 
0.9294& 
4.574& 
3.650 \\
\hline
8.0& 
8.0& 
0.9884& 
4.577& 
3.651 \\
\hline
10.0& 
10.0& 
0.9957& 
4.580& 
3.653 \\
\hline
12.0& 
12.0& 
0.9982& 
4.583& 
3.655 \\
\hline
\end{tabular}
\end{center}
\caption{Degree of entanglement for different values of waveguide depths and widths with the corresponding values of $\Lambda_{1}$ and $\Lambda_{2}$.}
\label{Table_2}
\end{table}

\newpage

\bibliography{Thyagarajan_Polar_entanglement_resub_PRA}

\end{document}